\documentclass[aps, pre, reprint, showpacs, showkeys]{revtex4-1}
\usepackage{graphicx}
\begin{document}
\newcommand{\eqlabel}[1]{\label{eq:#1}}
\newcommand{\eqref}[1]{Eq.~(\ref{eq:#1})}
\newcommand{\figlabel}[1]{\label{fig:#1}}
\newcommand{\figref}[1]{Fig.~\ref{fig:#1}}
\newcommand{\secref}[1]{Sec.~\ref{sec:#1}}
\newcommand{\etal}{et al. }
\newcommand{\apriori}{\textit{a priori~}}
\newcommand{\scf}{\Psi}
\title{Bayesian Inference in the Scaling Analysis of Critical Phenomena}
\author{Kenji Harada}
\affiliation{Graduate School of Informatics,
  Kyoto University, Kyoto 606-8501, Japan}
\begin{abstract}
  To determine the universality class of critical phenomena, we
  propose a method of statistical inference in the scaling analysis of
  critical phenomena. The method is based on Bayesian statistics, most
  specifically, the Gaussian process regression. It assumes only the
  smoothness of a scaling function, and it does not need a form. We
  demonstrate this method for the finite-size scaling analysis of the
  Ising models on square and triangular lattices. Near the critical
  point, the method is comparable in accuracy to the least-square
  method. In addition, it works well for data to which we cannot apply
  the least-square method with a polynomial of low degree. By
  comparing the data on triangular lattices with the scaling function
  inferred from the data on square lattices, we confirm the
  universality of the finite-size scaling function of the
  two-dimensional Ising model.
\end{abstract}
\maketitle

%
%
\section{Introduction}
\label{sec:intro}
A wide variety of systems exhibit critical phenomena. Near a critical
point, some quantities obey scaling laws. As an example, consider
\begin{equation}
  A(t, h) = t^x\scf(ht^{-y}),  
  \eqlabel{s_f}
\end{equation}
where $t$ and $h$ are variables describing a system, and the critical
point is located at $t=h=0$. The scaling law is derived by the
renormalization group
argument\cite{goldenfeld92:_lectur_phase_trans_and_renor_group,
  *cardy96:_scalin_and_renor_in_statis_physic}.  The scaling exponents
$x$ and $y$ are called \textit{critical exponents}. The universality
of critical phenomena means that different systems share the same set
of critical exponents. Thus, this set defines a \textit{universality
  class} of critical phenomena. In addition, the \textit{scaling
  function} $\scf$ also exhibits universality. For example, Mangazeev
\etal numerically obtained scaling functions of the Ising models on
square and triangular lattices\cite{2010PhRvE..81f0103M}. Since Ising
models on both lattices belong to the same universality class, the two
scaling functions with nonuniversal metric factors are perfectly
equal.

An important issue to study critical phenomena is to determine the
universality class. The object of \textit{scaling analysis} is to
determine the universality class from data. We assume the scaling law
of \eqref{s_f} for data. If we plot data with rescaled coordinates as
$(X_i, Y_i) \equiv (h_it_i^{-y}, t_i^{-x}A(t_i, h_i))$, all points
must collapse on a scaling function as $Y_i=\scf(X_i) $. To determine
critical exponents, we need a mathematical method to estimate how well
all rescaled points collapse on a function for a given set. In other
words, we need to estimate the goodness of data
collapse. Unfortunately, we do not know the form of $\scf$
\apriori. The conventional method for the scaling analysis is a
least-square method while assuming a polynomial.  However, it may be
difficult to choose the degree of the polynomial for data, because
there are overfitting problems associated with increasing the
degree. To use a polynomial of low degree, we usually limit the data
to a narrow region near a critical point.  However, it may require
high accuracy. In addition, it may be difficult to obtain a universal
scaling function in a wide critical region. Thus, the scaling analysis
by the least-square method must be carefully done as shown in the
reference \cite{slevin99:_correc_to_scalin_at_ander_trans}.

In this paper, we propose a method of statistical inference in the
scaling analysis of critical phenomena. The method is based on
Bayesian statistics.  Bayesian statistics has been widely used for
data analysis \cite{bishop06:_patter_recog_and_machin_learn}. However,
to the best of our knowledge, it has not been applied to the scaling
analysis of critical phenomena. In particular, since our method
assumes only the smoothness of a scaling function, it can be applied
to data for which the least-square method cannot be used.

In \secref{method}, we first introduce a Bayesian framework in the
scaling analysis of critical phenomena. Next, we propose a Bayesian
inference using a Gaussian process (GP) in this framework. In
\secref{demo}, we demonstrate this method for critical phenomena of
the Ising models on square and triangular lattices. Finally, we
give the conclusions in \secref{conclusion}.

%
%

\section{Bayesian framework and Bayesian inference in scaling
  analysis}
\label{sec:method}
By using two functions $X$ and $Y$ that calculate rescaled
coordinates, the scaling law of an observable $A$ can be rewritten as
\begin{equation}
  \label{eq:law}
  Y(A(\vec{v}), \vec{v}, \vec{\theta_p}) = \scf(X(\vec{v}, \vec{\theta_p})),
\end{equation}
where $\vec{v}$ denotes the variables describing a system and
$\vec{\theta_p}$ denotes the additional parameters as critical
exponents.  Our purpose is to infer $\vec{\theta_p}$ so that data
$A(\vec{v_i}), (1 \le i \le N)$ obey the scaling law of \eqref{law}.
In the following, for convenience, we abbreviate $X(\vec{v_i},
\vec{\theta_p})$ and $Y(A(\vec{v_i}), \vec{v_i}, \vec{\theta_p})$ to
$X_i$ and $Y_i$, respectively.

When the statistical error of $Y_i$ is $E_i$, the distribution
function of $\{Y_i\}$, $P(\vec{Y}|\scf, \vec{\theta_p})$, is a
multivariate Gaussian distribution with mean vector $\vec{\scf}$ and
covariance matrix $\mathcal{E}$:
\begin{equation}
  P(\vec{Y}|\scf,\vec{\theta_p}) \equiv 
  \mathcal{N}(\vec{Y}|\vec{\scf}, \mathcal{E}),
  \eqlabel{pofp}
\end{equation}
where $(\vec{Y})_i \equiv Y_i$, $(\vec{\scf})_i \equiv \scf(X_i)$,
$(\mathcal{E})_{ij}\equiv E_i^2\delta_{ij} $, and
$$
\mathcal{N}(\vec{y}|\vec{\mu}, \Sigma) \equiv
\frac{1}{\sqrt{|2\pi\Sigma|}} \exp\left(-\frac12
  (\vec{y}-\vec{\mu})^{t} \Sigma^{-1} (\vec{y}-\vec{\mu}) \right) .
$$

Next, we introduce a statistical model for a scaling function as
$P(\scf | \vec{\theta_h})$.  Here, $\vec{\theta_h}$ denotes the
control parameters and is referred to as \textit{hyper parameters}.
Then, the conditional probability of $\vec{Y}$ for $\vec{\theta_p}$
and $\vec{\theta_h}$ is formally defined as
\begin{eqnarray}
  P(\vec{Y} | \vec{\theta_p}, \vec{\theta_h})
  \equiv \int   P(\vec{Y}|\scf, \vec{\theta_p})
  P(\scf | \vec{\theta_h}) d\scf.
  \eqlabel{conditional}
\end{eqnarray}
According to Bayes' theorem, a conditional probability of
$\vec{\theta_p}$ and $\vec{\theta_h}$ for $\vec{Y}$ can be written as
\begin{equation}
  P(\vec{\theta_p},\vec{\theta_h} | \vec{Y}) =
  {P(\vec{Y} | \vec{\theta_p}, \vec{\theta_h}) P(\vec{\theta_p},
    \vec{\theta_h})}/ {P(\vec{Y})},
  \eqlabel{bayes-th}
\end{equation}
where $P(\vec{\theta_p}, \vec{\theta_h})$ and $P(\vec{Y})$ denote the
\textit{prior distributions} of $\vec{\theta_p}$ and $\vec{\theta_h}$
and that of $\vec{Y}$, respectively. In Bayesian statistics,
$P(\vec{\theta_p}, \vec{\theta_h} | \vec{Y})$ is called a
\textit{posterior distribution} of $\vec{\theta_p}$ and
$\vec{\theta_h}$. Using \eqref{bayes-th}, a posterior probability of
$\vec{\theta_p}$ and $\vec{\theta_h}$ for $\vec{Y}$ can be
estimated. This is a Bayesian framework for the scaling analysis of
critical phenomena.

In Bayesian statistics, the conventional method of inferring
parameters is the maximum a posteriori (MAP) estimate.  In this paper,
for simplicity, we assume that all prior distributions are
uniform. Then,
\begin{equation}
    P(\vec{\theta_p},\vec{\theta_h} | \vec{Y}) 
    \propto P(\vec{Y} | \vec{\theta_p}, \vec{\theta_h}).
    \eqlabel{bayes_th}
\end{equation}
Therefore, the MAP estimate is equal to a maximum
\textit{likelihood}(ML) estimate with a likelihood function of
$\vec{\theta_p}$ and $\vec{\theta_h}$, defined as
\begin{equation}
  \eqlabel{likelihood}
\mathcal{L}(\vec{\theta_p}, \vec{\theta_h}) =
P(\vec{Y}| \vec{\theta_p}, \vec{\theta_h}).
\end{equation}
In addition, the confidence intervals of the parameters can be
estimated through \eqref{bayes_th}.

In this framework, the statistical model of a scaling function plays
an important role. We start from a polynomial scaling function as
$\scf(X) \equiv \sum_{k} c_k X^k$. If a coefficient $c_k$ is
distributed by a probability density $P(c_k|\vec{\theta_h})$, then
$P(\scf|\vec{\theta_h})d\scf \equiv \prod_k
P(c_k|\vec{\theta_h})dc_k$. We first consider the strong constraint
for $c_k$ as $ P(c_k|\vec{\theta_h}) \equiv \delta(c_k-m_k)$, where
$m_k$ is a hyper parameter.  Then, $P(\vec{Y}|\vec{\theta_p},
\vec{\theta_h})$ is a multivariate Gaussian distribution with mean
vector $\vec{\mu}$ and covariance matrix $\Sigma$:
\begin{equation}
  \label{eq:mean-vector}
  (\vec{\mu})_i \equiv \sum_k m_k {X_i}^k, \quad \Sigma \equiv \mathcal{E}.
\end{equation}
Thus, the ML estimate in \eqref{likelihood} is equal to the
least-square method. We soften this constraint as
$P(c_k|\vec{\theta_h})\equiv\mathcal{N}(c_k|m_k, \sigma_k^2)$, where
$m_k$ and $\sigma_k$ are hyper parameters. Then,
$P(\vec{Y}|\vec{\theta_p}, \vec{\theta_h})$ is again a multivariate
Gaussian distribution, and the covariance matrix changes as follows:
\begin{equation}
\Sigma \equiv \mathcal{E} + \Sigma', \quad (\Sigma')_{ij} \equiv \sum_k
(X_i X_j)^k \sigma_k^2.
\eqlabel{sigma-poly}
\end{equation}
This includes the case of a strong constraint such as $\sigma_k^2 =
0$.

To calculate a MAP estimate, a log-likelihood function is used. If a
posterior distribution is described by a multivariate Gaussian
function as $P(\vec{\theta_p}, \vec{\theta_h} | \vec{Y}) \propto
\mathcal{N}(\vec{Y}|\vec{\mu},\Sigma)$, the log-likelihood function
can be written as
\begin{equation}
  \log \mathcal{L}(\vec{\theta_p}, \vec{\theta_h}) \equiv
  -\frac{1}{2} \log\left|2\pi\Sigma\right|
  -\frac12 (\vec{Y}-\vec{\mu})^t \Sigma^{-1}(\vec{Y}-\vec{\mu}).
    \eqlabel{BayesGP0}
\end{equation}
Although the likelihood function is nonlinear in parameters
$\vec{\theta_p}$ and $\vec{\theta_h}$, a multidimensional maximization
method may be applied to calculate a MAP estimate. Under a strong
constraint such as $\sigma_k^2 = 0$, the Levenberg-Marquardt algorithm
is efficient. Under a weak constraint such as $\sigma_k^2 > 0$, we may
use an efficient maximization algorithm such as the Fletcher-Reeves
conjugate gradient algorithm. In such efficient algorithms, we
sometimes need the derivative of \eqref{BayesGP0} for a parameter
$\theta$. Then, we can use the following formula:
\begin{eqnarray}
\frac{\partial \log \mathcal{L}(\vec{\theta_p}, \vec{\theta_h})}{
    \partial \theta} &=& -\frac{1}{2}\mathbf{Tr}\left(\Sigma^{-1}
\frac{\partial \Sigma}{\partial \theta}\right)\nonumber\\
&-& (\vec{Y}-\vec{\mu})^t \Sigma^{-1}\frac{\partial
  (\vec{Y}-\vec{\mu})}{\partial \theta}
\nonumber\\
&+& \frac{1}{2}(\vec{Y}-\vec{\mu})^t \Sigma^{-1}
\frac{\partial\Sigma}{\partial \theta}  \Sigma^{-1}
(\vec{Y}-\vec{\mu}).
\label{eq:derivative}
\end{eqnarray}
However, to compute the inverse of a covariance matrix, the
computational cost of an iteration is $O(N^3)$. On the other hand,
$O(N^2)$ for the least-square method. Fortunately, using a
high-performance numerical library for linear algebra, we can easily
make a code and we can efficiently calculate for some hundred data
points. Another method is based on Monte Carlo (MC) samplings. In
particular, MC samplings may be useful for the estimate of the
confidence intervals of parameters.

We demonstrate the MAP estimate based on \eqref{BayesGP0} and
\eqref{sigma-poly}. \figref{test-sub1} shows the data points rescaled
by a MAP estimate. Here, we assume that a scaling function is
linear. To show the flexibility of Bayesian inference, we fix
$m_0=m_1=0$.  Thus, $\sigma_0$ and $\sigma_1$ are the only free
parameters. We artificially generate mock data so that they obey a
scaling law:
\begin{equation}
  \label{eq:fss}
A(T,L)=L^{-\beta/\nu}\scf((T-T_c)L^{1/\nu}),
\end{equation}
where $T$ and $L$ denote the temperature and linear dimension of a
system, respectively.  This is a well-known scaling law for
finite-size systems. In \figref{test-sub1}, we set $T_c=\beta/\nu=1,
1/\nu=2 $ and $\scf(X)=2+X$. Then,
\begin{equation}
  \eqlabel{mock}
  A_i = \frac{2}{L_i} + (T_i-1)L_i + r_i / 50,
\end{equation}
where $r_i$ is a Gaussian noise. These mock data are shown in the
inset of the left panel of \figref{test-sub1}. The right panel of
\figref{test-sub1} shows the maximization of a likelihood, when we
start from $T_c=1/\nu=\beta/\nu=0$ and $\sigma_0=\sigma_1=2$.  The
results for $T_c, \beta/\nu$, and $1/\nu$ are $1.00745$, $0.999008$,
and $2.00638$, respectively. They are close to the correct values.

\begin{figure}
  {\includegraphics[width=0.25\textwidth]{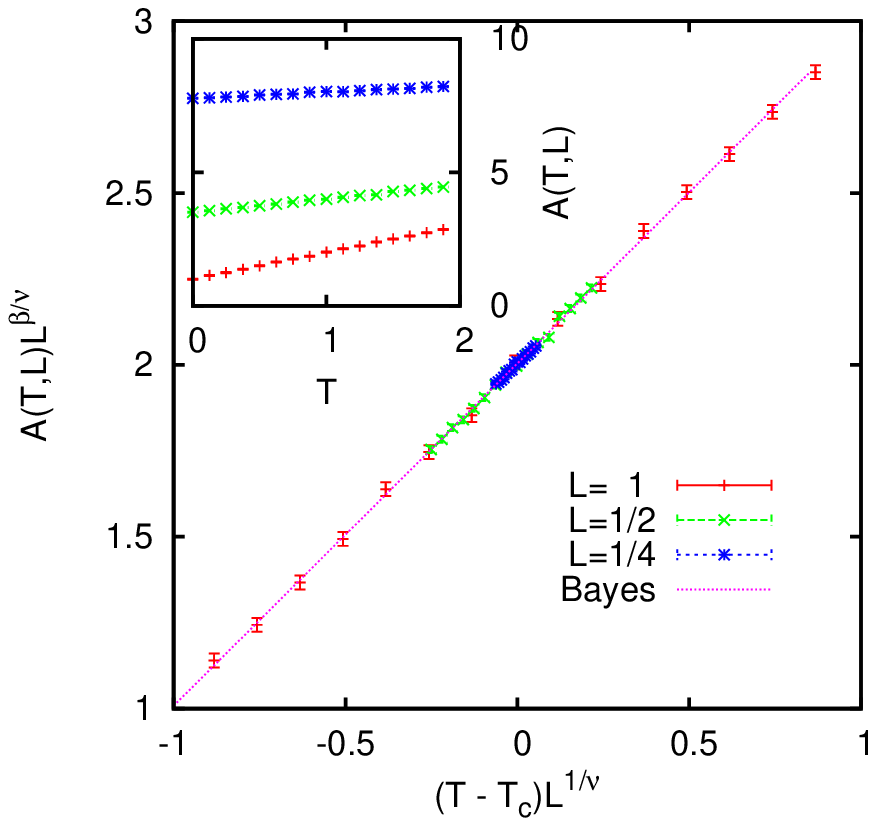}}
  {\includegraphics[width=0.225\textwidth]{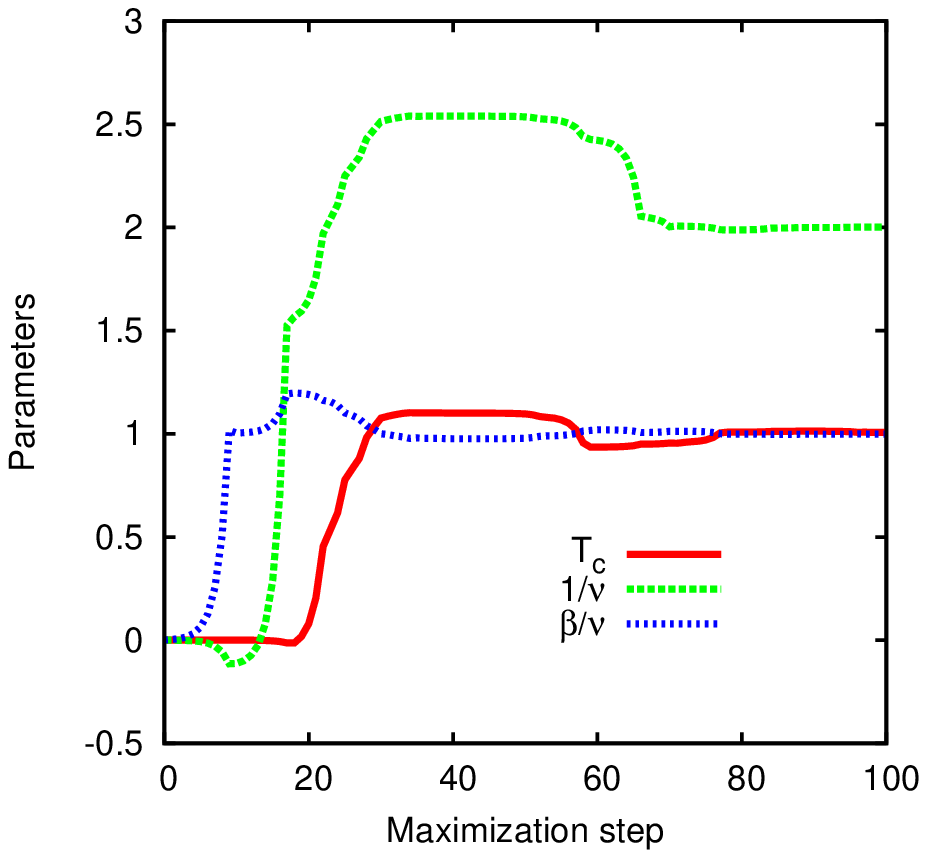}}
  \caption{(Color on-line) Left panel: 
    The data points rescaled by a MAP estimate. We assume that a scaling
    function is linear. The results of the MAP estimate for $T_c$,
    $\beta/\nu$, and $1/\nu$ are $1.00745$, $0.999008$, and $2.00638$, respectively.
    The dotted (pink) line is the scaling function inferred from the
    MAP estimate. Inset of left panel: Mock data set.  Right panel:
    Maximization of a likelihood.  \figlabel{test-sub1}}
\end{figure}

Unfortunately, we usually do not know the form of a scaling function
\apriori. The Bayesian inference based on \eqref{BayesGP0} and
\eqref{sigma-poly} may not be effective in some cases. Thus, we
consider an extension of \eqref{sigma-poly}. From \eqref{BayesGP0}, we
may regard data points as obeying a GP. Since the covariance matrix
represents statistical correlations in data, we may design it for a
wide class of scaling functions. Thus, we introduce a generalized
covariance matrix $\Sigma$ as
\begin{equation}
  \Sigma = \mathcal{E} + \Sigma', (\Sigma')_{ij} \equiv K(X_i, X_j),
  \label{eq:covariance}
\end{equation}
where $K(X_i, X_j)$ is called a \textit{kernel function}. Note that
$\Sigma'$ must be a positive definite. The Bayesian inference based on
\eqref{BayesGP0} and \eqref{covariance} is called a \textit{GP
  regression}. \eqref{sigma-poly} is a special case of
\eqref{covariance}. As shown in \figref{test-sub1}, even if
$\vec{\mu}=0$, the GP regression is successful. For simplicity, we
consider only a zero mean vector ($\vec{\mu}=0$) in this paper.

In the GP regression, we can also infer the scaling function. In fact,
we assume that all data points obey a GP. In other words, the joint
probability distribution of obtained data points and a new additional
point $(X, Y)$ is also a multivariate Gaussian
distribution. Therefore, a conditional probability of $Y$ for obtained
data can be written by a Gaussian distribution with mean $\mu(X)$ and
variance $\sigma^{2}(X)$:
\begin{equation}
  \eqlabel{meanGPK}
  \mu(X) \equiv
  \vec{k}^t\Sigma^{-1}\vec{Y},\quad
  \sigma^{2}(X)
  \equiv K(X, X)-\vec{k}^t\Sigma^{-1}\vec{k},
\end{equation}
where $(\vec{k})_{i} \equiv K(X_{i}, X)$. We regard $\mu(X)$ in
\eqref{meanGPK} as a scaling function. For example, the dotted (pink)
line in \figref{test-sub1} is $\mu(X)$ in \eqref{meanGPK} for mock
data with a MAP estimate.

In general, a scaling function is smooth. Since $\mu(X)$ in
\eqref{meanGPK} is the weighted sum of kernel functions, the kernel
function should smoothly decrease for increasing distance between two
arguments. In this paper, we propose the use of a \textit{Gaussian
  kernel function} (GKF) for the scaling analysis of critical
phenomena. GKF is defined as
\begin{equation}
  \label{eq:GKF}
  K_G(X_i, X_j) \equiv \theta_0^2 \exp \left( -\frac{(X_i - X_j)^2}{2\theta_1^2}
  \right), 
\end{equation}
where $\theta_0$ and $\theta_1$ are hyper parameters. Since GKF is
smooth and local, the GP regression with GKF may be effective for a
wide class of scaling functions.

%
%
\section{Bayesian finite-size scaling analysis of the two-dimensional
  Ising model}
\label{sec:demo}
We demonstrate the GP regression with GKF for the finite-size scaling
(FSS) analysis of the two-dimensional Ising model. FSS is widely used
in numerical studies of critical phenomena for finite-size systems.
It is based on the FSS law derived by the renormalization group
argument. The Hamiltonian of the Ising model can be written as
\begin{equation}
  \label{eq:Ising}
  \mathcal{H}(\{s_i\}) \equiv - J \sum_{\langle ij \rangle} s_i s_j,
\end{equation}
where $s_i$ is the spin variable ($\pm 1$) of site $i$ and $\langle ij
\rangle$ denotes the nearest neighbor pairs and $J$ denotes a positive
coupling constant. The partition function can be written as
\begin{equation}
  \label{eq:partition}
  Z \equiv \sum_{\{s_i\}} \exp[-H(\{s_i\})/k_BT],
\end{equation}
where $k_B$ is the Boltzmann constant. For simplicity, we set
$J/k_B=1$ in the following. The two-dimensional Ising model has a
continuous phase transition at a finite temperature. Since there are
exact results for the Ising models on square and triangular
lattices\cite{onsager44:_cryst_statis, *1952PhRv...85..808Y}, we can
check the results of FSS.
\begin{figure}
  \includegraphics[width=0.5\textwidth]{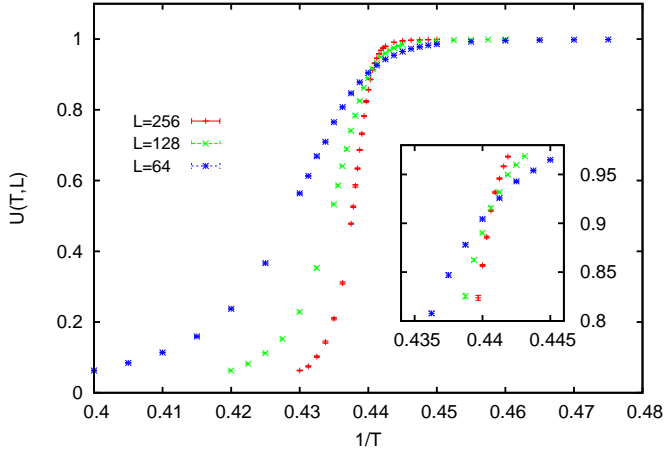}
  \caption{\figlabel{Binder-ratio} (Color on-line) The Binder ratios of
    the Ising model on three square lattices. The total number of data items
    is 86. Inset: Binder ratio near a critical point. The value of the
    Binder ratio is limited to the region $[0.8, 0.97]$. The number of
    data items in the inset is $24$.  }
\end{figure}
To obtain the Binder ratios\cite{1981ZPhyB..43..119B} and magnetic
susceptibility on square and triangular lattices, MC simulations
have been done.　For the square lattice, $L=L_r=64$, $128$, and $256$,
where $L_r$ and $L$ denote the number of rows and columns of the
lattice, respectively. For the triangular lattice, $L=(65L_r/75)=65$,
$130$, and $260$ so that the aspect ratio of a triangular lattice is
approximately $1$. We set periodic boundary conditions for both
lattices. The number of MC sweeps by the cluster
algorithm\cite{swendsen87:_nonun_critic_dynam_in_monte_carlo_simul} is
$80000$ for each simulation. The Binder ratio is based on the ratio of
the fourth and second moments of an order parameter. The order
parameter of the Ising model is a magnetization defined as $M \equiv
\sum_i s_i$. Then, the Binder ratio can be written as
\begin{equation}
  \label{eq:def-binder}
  U \equiv \frac{1}{2}\left(3 - \frac{\langle M^4\rangle}{\langle
M^2\rangle^2}\right),
\end{equation}
where $\langle \cdot \rangle$ denotes the canonical ensemble average.
In the thermodynamic limit, the Binder ratio takes values 1 and 0 in
the order and disorder phases, respectively. Since the Binder ratio is
dimensionless, the FSS form is
\begin{equation}
  \label{eq:scaling-form-B}
U(T, L) = \scf_B ((1/T -1/T_c)L^{1/\nu}),  
\end{equation}
where $T_c$ is a critical temperature and $\nu$ is a critical exponent
that characterizes the divergence of a magnetic correlation length.
From \eqref{scaling-form-B}, the value of the Binder ratio at the
critical temperature is universal. Magnetic susceptibility can be
written as
\begin{equation}
  \label{eq:sus}
  \chi \equiv \frac{1}{TV} \left(\langle M^2 \rangle - \langle M \rangle^2\right),
\end{equation}
where $V$ is the number of spins. The scaling form of magnetic
susceptibility is
\begin{equation}
  \label{eq:sf-sus}
  \chi(T,L) = L^{\gamma/\nu} \scf_\chi((1/T - 1/T_c)L^{1/\nu}),
\end{equation}
where $\gamma$ and $\nu$ are critical exponents.

\begin{figure}
  \includegraphics[width=0.5\textwidth]{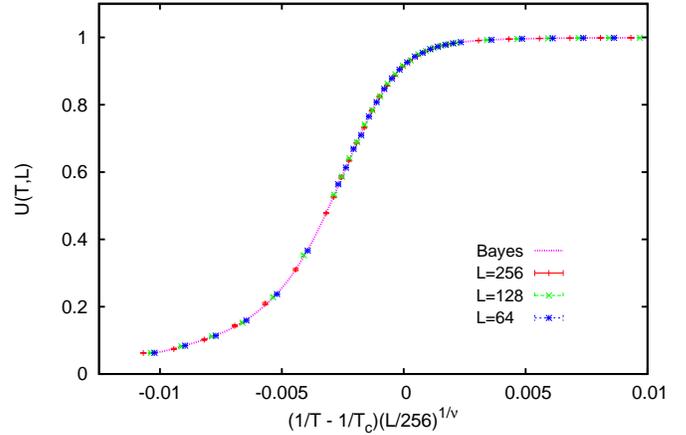}
  \caption{\figlabel{Bayes-S-B} (Color on-line) Result of a Bayesian FSS
    of the Binder ratio of the Ising model on square lattices. We
    apply the GP regression with \eqref{kernel} to the data shown in
    \figref{Binder-ratio}. The results of the MC estimate are
    $1/T_c=0.440683(7)$ and $1/\nu = 0.996(2)$. The dotted (pink) curve is
    the scaling function inferred from a MAP estimate.  }
\end{figure}
\begin{figure}
  \includegraphics[width=0.5\textwidth]{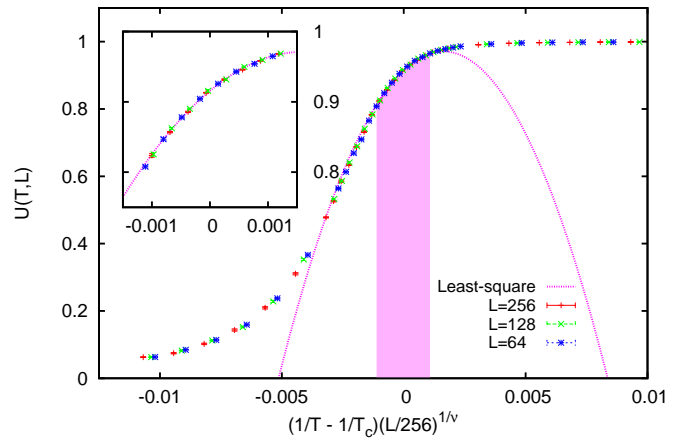}
  \caption{\figlabel{CS-BR} (Color on-line) Result of a FSS of the
    Binder ratio of the Ising model on square lattices by the
    least-square method. In the least-square method, we use only the data in the
    inset of \figref{Binder-ratio} and assume that the scaling
    function is a quadratic function. The best estimate of the
    least-square method is $1/T_c=0.44069(2)$ and
    $1/\nu=1.00(2)$. All data points are rescaled by these values.
    The data points used in the least-square method are shown in the filled
    gray (pink) region. The dotted (pink) curve is the scaling function
    inferred from the best estimate of the least-square method. Inset:
    Rescaled data points used in the least-square method.  }
\end{figure}
\begin{figure}
  \includegraphics[width=0.235\textwidth]{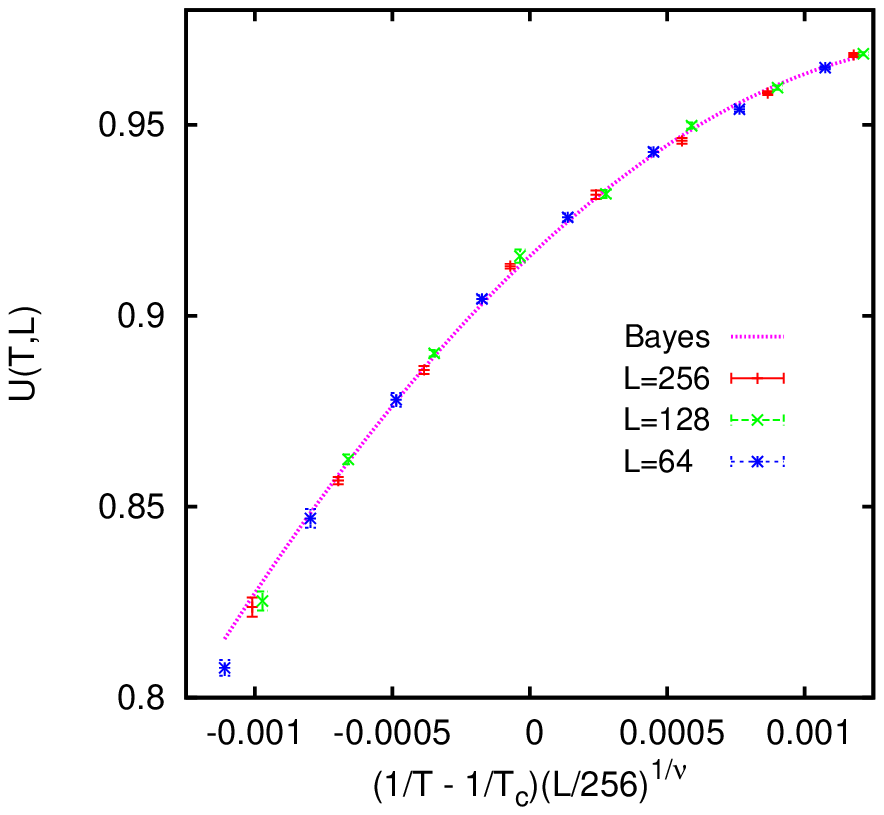}
  \includegraphics[width=0.235\textwidth]{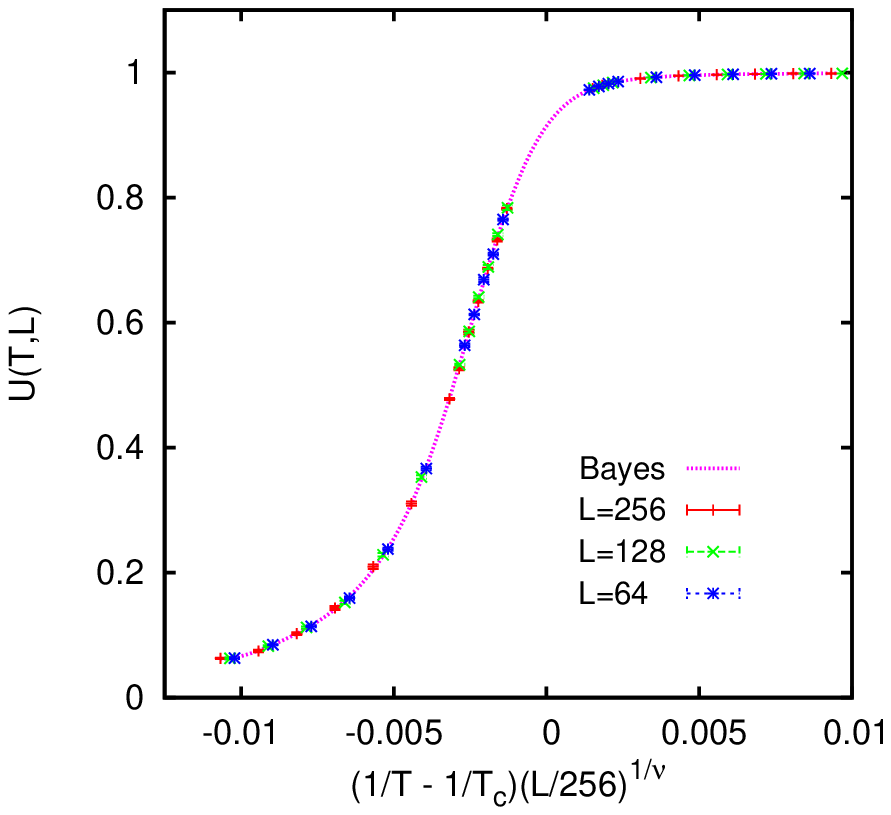}
  \caption{\figlabel{Bayes2-S-B} (Color on-line) Left panel: Result of
    a Bayesian FSS for the data in the inset of
    \figref{Binder-ratio}. The results of the MC estimate are
    $1/T_c=0.44070(2)$ and $1/\nu = 1.00(1)$.  Right panel: Result of
    a Bayesian FSS for data not included in the inset of
    \figref{Binder-ratio}. The results of the MC estimate are
    $1/T_c=0.440675(9)$ and $1/\nu = 0.997(2)$.  The dotted (pink)
    curves in left and right panels are the scaling functions inferred
    from the MAP estimates.  }
\end{figure}
We first apply the GP regression to the Binder ratios of square
lattices shown in \figref{Binder-ratio}. The kernel function based on
GKF can be written as
\begin{equation}
  \eqlabel{kernel}
  K(X_i, X_j) \equiv K_G(X_i, X_J) + \theta_2^2 \delta_{ij},
\end{equation}
where a hyper parameter $\theta_2$ denotes the data fidelity. We note
that the maximization of a likelihood is much improved by
$\theta_2$. Although $\theta_2$ finally goes to zero, it helps to
escape from a local maximum of a likelihood. \figref{Bayes-S-B} shows
the result of the GP regression for Binder ratios. The results of the
MC estimate are $1/T_c=0.440683(7)$ and $1/\nu = 0.996(2)$. This is
consistent with the exact results $1/T_c =
\ln(1+\sqrt{2})/2=0.4406867925\cdots$ and $1/\nu=1$.  The dotted
(pink) curve in \figref{Bayes-S-B} is the scaling function inferred
from a MAP estimate by using \eqref{meanGPK}. All points collapse on
this curve. The value of the Binder ratio at the critical temperature
is $0.9158(4)$. This is consistent with the exact value
$0.916038\cdots$\cite{Salas:2000fk}. It is difficult to represent this
curve as a polynomial of low degree. Thus, we limit the value of a
Binder ratio to the region $[0.8, 0.97]$ (see the inset of
\figref{Binder-ratio}). We apply the least-square method with a
quadratic function to the limited data.  The result is shown in
\figref{CS-BR}. The inset of \figref{CS-BR} shows the data points
rescaled by the best estimate of the least-square method. All points
in the inset collapse on a quadratic function (see the dotted (pink)
curve in \figref{CS-BR}). The reduced chi-square is $2.96$. The
results of the least-square method are $1/T_c = 0.44069(2)$ and $1/\nu
=1.00(2)$. This is consistent with the exact result. However, it may
be difficult to extend the region of data for the least-square
method. The main panel of \figref{CS-BR} shows all data points
rescaled by the best estimate of the least-square method. While all
points again collapse on a smooth curve, the curve is not equal to the
quadratic function outside the limited region (see the filled gray
(pink) region in \figref{CS-BR}). The left panel in
\figref{Bayes2-S-B} shows the result of the GP regression to the same
data for the least-square method.  The results of the MC estimate are
$1/T_c=0.44070(2)$ and $1/\nu = 1.00(1)$. This is consistent with the
exact results and similar to that of the least-square method. The GP
regression with GKF assumes only the smoothness of a scaling
function. Thus, it may be effective even for the data not near a
critical point. In fact, even if we use only data not included in the
inset of \figref{Binder-ratio}, we can do FSS by the GP
regression. The result is shown in the right panel in
\figref{Bayes2-S-B}. The results of the MC estimate are
$1/T_c=0.440675(9)$ and $1/\nu=0.997(2)$. Although we do not use the
important data near a critical point, the result of the GP regression
is close to the exact result.

\begin{figure}
  \includegraphics[width=0.5\textwidth]{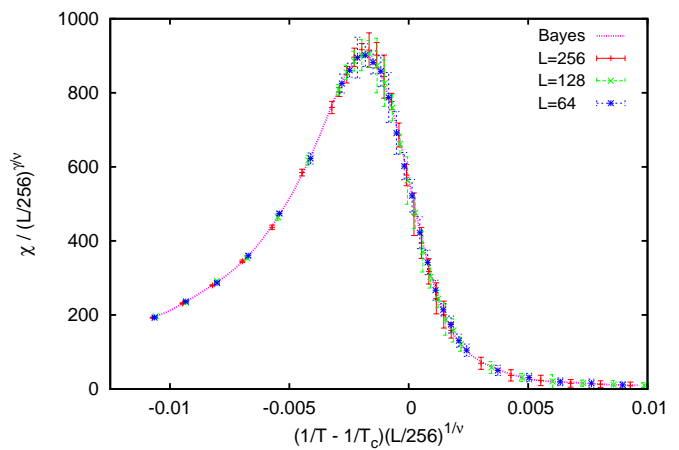}
  \caption{\figlabel{Bayes-S-X} (Color on-line) Result of a Bayesian FSS
    of the magnetic susceptibility of the Ising model on square
    lattices. We apply the GP regression with \eqref{kernel} to data
    with the same temperatures and lattice sizes of data as in
    \figref{Binder-ratio}. The results of the MC estimate are
    $1/T_c=0.44072(8)$, $1/\nu = 0.98(2)$, and $\gamma / \nu =
    1.74(2)$. The dotted (pink) curve is the scaling function inferred
    from a MAP estimate.  }
\end{figure}
We also apply the GP regression to the magnetic susceptibility of
square lattices. The result is shown in \figref{Bayes-S-X}. The
results of the MC estimate are $1/T_c=0.44072(8)$, $1/\nu=0.98(2)$,
and $\gamma/\nu=1.74(2)$. This is consistent with the exact result
($\gamma/\nu=7/4=1.75$). The dotted (pink) curve is the scaling
function inferred from the MAP estimate by using \eqref{meanGPK}.  All
points collapse on this curve. However, it is difficult to represent
this curve as a polynomial of low degree.

\begin{figure}
  \includegraphics[width=0.5\textwidth]{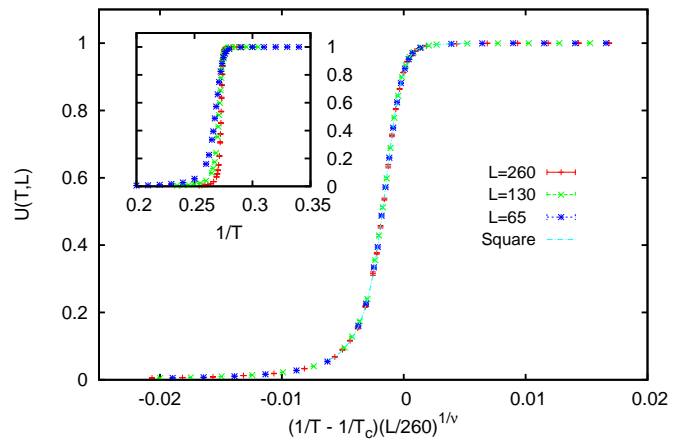}
  \caption{\figlabel{Bayes-T-B} (Color on-line) Result of a Bayesian
    FSS of the Binder ratio of the Ising model on triangular
    lattices. We apply the GP regression with \eqref{kernel}.  The
    results of the MC estimate are $1/T_c=0.274652(7)$ and $1/\nu =
    0.989(4)$. The dashed (light-blue) curve is the scaling function
    of a square lattice in \figref{Bayes-S-B} with a nonuniversal
    metric factor $C_1=1.748(3)$. Inset: Binder ratios of the Ising
    model on triangular lattices. The number of data items is 86. }
\end{figure}
\begin{figure}
  \includegraphics[width=0.5\textwidth]{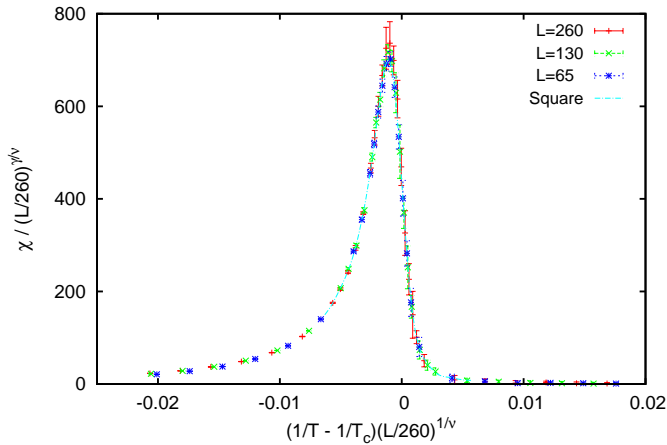}
  \caption{\figlabel{Bayes-T-X} (Color on-line) Result of a Bayesian
    FSS of the magnetic susceptibility of the Ising model on
    triangular lattices.  We apply the GP regression with
    \eqref{kernel} to the data with the same temperatures and lattice
    sizes of data as in \figref{Bayes-T-B}. The results of the MC
    estimate are $1/T_c=0.27466(7)$, $1/\nu = 0.95(2)$, and $\gamma /
    \nu = 1.71(2)$.  The dashed (light-blue) curve is the scaling
    function of a square lattice in \figref{Bayes-S-X} with
    nonuniversal metric factors $C_1=1.70(2)$ and $C_2=0.777(7)$.  }
\end{figure}
Next, we apply the GP regression to the Binder ratio and magnetic
susceptibility on triangular lattices. These results are shown in
\figref{Bayes-T-B} and \figref{Bayes-T-X}, respectively. All points of
each quantity collapse on a curve. The results of the MC estimate for
$1/T_c$, $1/\nu$, and $\gamma/\nu$ are summarized in
Tab.~\ref{tab:all-results}. Although they are almost consistent with
the exact results, the accuracy of inference is lower than that for
the data of square lattices. Since the region of the data of
triangular lattices is wide (compare \figref{Bayes-T-B} with
\figref{Bayes-S-B}), we may consider the correction to scaling.
\begin{table*}[ht]
  \label{tab:all-results}
  \centering
  \caption{Results of the MC estimates for $1/T_c$, $1/\nu$, and $\gamma/\nu$.
    The exact values of $1/T_c$ for square and triangular lattices
    \cite{onsager44:_cryst_statis, *1952PhRv...85..808Y} are
    $\ln(1+\sqrt{2})/2=0.4406867925\cdots$ and
    $(\ln3)/4=0.2746530723\cdots$, respectively. The exact values of
    $1/\nu$ and $\gamma/\nu$ are $1$ and $\frac74$, respectively.
  }
  \begin{ruledtabular}
  \begin{tabular}{|l|l|l|l|l|l|}
    Data & Lattice & Method & $1/T_c$ & $1/\nu$ & $\gamma/\nu$\\
    \hline
    Binder ratio & Square & GP regression & $0.440683(7)$ & $0.996(2)$ & -\\
    Binder ratio & Triangular & GP regression &
    $0.274652(7)$ & $0.989(4)$ & - \\
    Binder ratio\footnotemark[1] & Square & Least-square & $0.44069(2)$
    & $1.00(2)$ &-\\
    Binder ratio\footnotemark[1] & Square & GP regression & $0.44070(2)$ &
    $1.00(1)$ & -\\
    Binder ratio\footnotemark[2] & Square & GP regression &
    $0.440675(9)$ & $0.997(2)$ & -\\
    Magnetic susceptibility & Square & GP regression & $0.44072(8)$ & $0.98(2)$ & $1.74(2)$ \\
    Magnetic susceptibility & Triangular & GP regression & $0.27466(7)$ & $0.95(2)$ & $1.71(2)$\\
  \end{tabular}
  \footnotetext[1]{Data in the inset of \figref{Binder-ratio}.}
  \footnotetext[2]{Data not included in the inset of \figref{Binder-ratio}.}
\end{ruledtabular}
\end{table*}

Privman and Fisher proposed the universality of the finite-size
scaling function \cite{1984PhRvB..30..322P}. If two critical systems
belong to the same universality class, the two finite-size scaling
functions with nonuniversal metric factors are equal as
\begin{equation}
  \label{eq:UFSS}
  \scf(x) = C_2 \ \scf'(C_1x),
\end{equation}
where $\scf$ and $\scf'$ are finite-size scaling functions and $C_1$
and $C_2$ are nonuniversal metric factors. Hu \etal checked this idea
for bond and site percolation on various
lattices\cite{1995PhRvL..75..193H}.
The Ising models on square and triangular lattices belong to the
same universality. Thus, the two scaling functions must be equal via
nonuniversal metric factors as in \eqref{UFSS}. To check the
universality of finite-size scaling functions, we compared the data on
triangular lattices with the scaling function inferred from the data
on square lattices. We estimated nonuniversal metric factors to
minimize the residual between them. The result for the Binder ratio is
$C_1=1.748(3)$. The results for the magnetic susceptibility are
$C_1=1.70(2)$ and $C_2=0.777(7)$. Note that there is no metric factor
$C_2$ for the Binder ratio, because the Binder ratio is
dimensionless. The scaling functions of a square lattice with
nonuniversal metric factors are shown using the dashed (light-blue)
curves in \figref{Bayes-T-B} and \figref{Bayes-T-X}. They agree well
with the data on triangular lattices. The reduced chi-square of the
Binder ratio is $2.65$, and that of magnetic susceptibility is
$0.36$. Therefore, we confirm the universality of finite-size scaling
functions for the Binder ratio and magnetic susceptibility of the
two-dimensional Ising model. We note that Tomita \etal
\cite{Tomita:1999fk} confirmed the universality of finite-size scaling
functions for other quantities, and Mangazeev \etal
\cite{2010PhRvE..81f0103M} studied the universality of the scaling
function in the thermodynamic limit.

%
%
\section{Conclusions}
\label{sec:conclusion}

In this paper, we introduced a Bayesian framework in the scaling
analysis of critical phenomena. This framework includes the
least-square method for the scaling analysis as a special case. It can
be applied to a wide variety of scaling hypotheses, as shown in
\eqref{law}. In this framework, we proposed the GP regression with GKF
defined by Eqs. (\ref{eq:BayesGP0}), (\ref{eq:covariance}), and
(\ref{eq:GKF}). This method assumes only the smoothness of a scaling
function, and it does not need a form. We demonstrated it for the FSS
of the Ising models on square and triangular lattices.  For the data
limited to a narrow region near a critical point, the accuracy of the
GP regression was comparable to that of the least-square method. In
addition, for the data to which we cannot apply the least-square
method with a polynomial of low degree, our method worked
well. Therefore, we confirm the advantage of the GP regression with
GKF for the scaling analysis of critical phenomena.

The GP regression can also infer a scaling function as the mean
function $\mu$ in \eqref{meanGPK}. By comparing the data on triangular
lattices with the scaling function inferred from the data on square
lattices, we confirmed the universality of the FSS function of the
two-dimensional Ising model. The use of the scaling function may help
in the determination of a universality class.

In this paper, we assume that the data obey a scaling law. However, in
some cases, a part of the data may not obey the scaling law. In such a
case, we usually introduce a correction to scaling. If we can assume
the form of a correction to scaling, we change only the function $Y$
in \eqref{law}. However, the assumption of the correction term may
cause a problem. In other words, the identification of a critical
region remains.

As shown in this paper, the GP regression is a powerful method. In
particular, the GP regression can be applied to the statistical check
for data collapse.  For example, we can apply it to the estimate of
nonuniversal metric factors in \figref{Bayes-T-B} and
\figref{Bayes-T-X}. Another interesting application may be found in
the data analysis of physics.

\begin{acknowledgments}
  I would like to thank Toshio Aoyagi, Naoki Kawashima, Jie Lou, and
  Yusuke Tomita for the fruitful discussions, and the Kavli Institute
  for Theoretical Physics for the hospitality.  This research was
  supported in part by Grants-in-Aid for Scientific Research
  (No. 19740237, No. 22340111, No. 23540450), and in part by the
  National Science Foundation under Grant No. PHY05-51164.
\end{acknowledgments}

\bibliographystyle{apsrev4-1}
\bibliography{bayes}

\end{document}